\documentclass[%
 reprint,
 amsmath,amssymb,
 aps,
]{revtex4-2}
\usepackage{graphicx}% Include figure files
\usepackage{dcolumn}% Align table columns on decimal point
\usepackage{bm}
\usepackage{booktabs}
 \usepackage{multirow}
\usepackage{color,soul}
\usepackage{geometry}
\begin{document}
\preprint{APS/123-QED}
\title{Interface-resolved structural properties of epitaxial Y$_3$Fe$_5$O$_{12}$/ Gd$_3$Fe$_5$O$_{12}$ bilayers grown on GGG(111) by pulsed laser deposition}% Force line breaks with \\
\author{Kshitij Singh Rathore$^{1,2}$,  Abhisek Mishra$^{1,2}$,  Swayang Priya Mahanta$^{1,2}$, Shubhransu Sahoo$^{1,2}$, Anupama Swain$^{1,2}$, Pushpendra Gupta$^{1,2}$, Kapil Gupta$^{3}$,  José M. Caicedo-Roque$^{3}$, Jessica Padilla-Pantoja$^{3}$, Francisco J. Belarre $^{3}$, Belén Ballesteros$^{3}$, José Santiso$^{3\dagger}$, Subhankar Bedanta$^{1,2*}$}

\affiliation{$^{1}$Laboratory for Nanomagnetism and Magnetic Materials (LNMM), School of Physical Sciences, National Institute of Science Education and Research (NISER), Jatni, Odisha 752050, India}

\affiliation{$^{2}$Homi Bhabha National Institute, Training School Complex, Anushaktinagar, Mumbai 400094, India}

\affiliation{$^{3}$Catalan Institute of Nanoscience and Nanotechnology (ICN2), CSIC and BIST, Campus UAB, Bellaterra,
08193 Barcelona, Spain}

\thanks{Corresponding authors}
\email{\\ Email: sbedanta@niser.ac.in* \\  Email:  jose.santiso@icn2.cat$\dagger$}

\begin{abstract}
Epitaxial Y$_3$Fe$_5$O$_{12}$ (YIG) and Gd$_3$Fe$_5$O$_{12}$ (GdIG) thin films, along with their bilayer heterostructures, were grown on Gd$_3$Ga$_5$O$_{12}$ (GGG)(111) substrates using pulsed laser deposition. Structural properties were investigated using x-ray diffraction, reciprocal space mapping, and cross-sectional transmission electron microscopy. The results confirm high crystalline quality and coherent epitaxial growth, with RSM revealing a coexistence of strained and partially relaxed regions governed by layer sequence. TEM analysis shows sharp interfaces, columnar microstructures, and antiphase boundaries that facilitate strain relaxation. A comparative study indicates that the GGG/YIG/GdIG stacking sequence exhibits improved structural quality with reduced defect density, attributed to the superior epitaxial growth of YIG on the substrate. These findings highlight the critical role of growth sequence in controlling strain and interfacial structure in garnet heterostructures.
\end{abstract}

\maketitle
\section{Introduction}

Ferrimagnetic iron garnets constitute a central materials platform for insulating spintronics and magnonics due to their exceptionally low magnetic damping, long magnon propagation lengths, and compatibility with oxide heterostructures  \cite{Cherepanov1993,Serga2010,Chumak2015}. Among them, yttrium iron garnet (Y$_3$Fe$_5$O$_{12}$, YIG) has emerged as the prototypical system, serving as a benchmark material for studies of spin-wave transport, interfacial exchange phenomena, and coherent magnetization dynamics \cite{Hauser2016,Wang2014}.The renewed interest in YIG thin films has been driven by advances in thin-film growth techniques that enable the realization of epitaxial garnet layers with crystalline quality approaching that of bulk single crystals. In this context, rare-earth iron garnets further broaden the functional landscape of garnet-based oxide systems by introducing a magnetic rare-earth sublattice that is antiferromagnetically coupled to the dominant magnetisation of the Fe sublattices. Among them, gadolinium iron garnet (Gd$_3$Fe$_5$O$_{12}$, GdIG) has attracted considerable attention owing to its distinct lattice configuration and its sensitivity to epitaxial strain, growth conditions, and interfacial structural modifications, making it an effective model system for exploring interface-driven phenomena in garnet heterostructures\cite{Pauthenet1958,Roos2022}.
\begin{table}[htbp]
\centering
\caption{ Reported bulk crystallographic parameters such as lattice parameter($a$) and interplanar spacing ($d_{hkl}$) of GGG, YIG and GdIG are listed below. The lattice parameters are consistent with the ICDD Powder Diffraction File (GGG: PDF 01-083-4878, YIG: PDF 01-088-8938, and GdIG: PDF 01-072-0141) and the crystallographic data reported by Geller and Gilleo} \cite{Geller1963}.
\label{tab:lattice}
\begin{tabular}{lccc}
\hline
Material & $a$ (\AA) & $d_{444}$ (\AA) & $d_{2\bar20}$ (\AA)\\
\hline
GGG  & 12.379 & 1.7866 & 4.3766 \\
YIG  & 12.376--12.378 & 1.7864 & 4.3762 \\
GdIG & 12.470 & 1.7997 & 4.4093 \\
\hline
\end{tabular}
\end{table}

Heterostructures combining YIG and GdIG provide a unique all-insulating platform in which two ferrimagnetic garnets with distinct sublattice configurations are exchange-coupled across an epitaxial interface. The structural compatibility between YIG, GdIG, and GGG plays a crucial role in determining the epitaxial growth, strain state, and interface quality of the resulting heterostructures. Table~\ref{tab:lattice} summarizes the reported bulk cubic lattice parameters together with the corresponding interplanar spacings of the GGG substrate along with YIG and GdIG. 
Such YIG/GdIG bilayers can be viewed as garnet-based analogues of synthetic antiferromagnets \cite{gomez2018synthetic}, enabling unconventional magnetic states without introducing metallic layers or charge-current-related dissipation \cite{gomez2018synthetic, Becker2021}. These systems are therefore of interest both from a fundamental perspective and for future low-power magnonic architectures. These kind of heterostructures are very well suited and have been studied for spin dynamics \cite{fuhrmann2025temperature,kumar2022damping,zhuang2025effect,swain2024enhanced}, superconductivity \cite{reyren2007superconducting}, magnetoelastic coupling \cite{sgarro2025frequency}, two-dimensional electron
gases \cite{ohtomo2004high}, magnon-magnon coupling \cite{li2024reconfigurable,liu2024strong,rathore2026magnon},interfacial magnetism \cite{kim2025impact,keunecke2020high}, and exchange bias \cite{ghising2020spin,kumar2021positive,popova2003exchange}.

Despite their conceptual appeal, the growth of high-quality YIG/GdIG bilayers remains experimentally challenging. Although YIG and GdIG share the same garnet crystal structure, the substitution of non-magnetic Y$^{3+}$ ions by magnetic Gd$^{3+}$ ions profoundly alters the magnetic sublattice configuration and exchange pathways. As a result, the properties of YIG/GdIG heterostructures are extremely sensitive to deposition temperature, oxygen stoichiometry, post-deposition annealing, and interface abruptness \cite{Jakubisova2016,Mitra2017}. Even subtle interdiffusion or off-stoichiometry at the interface can lead to the formation of distinct interlayers, strongly modifying the net coercive behavior.

%This sensitivity has been clearly demonstrated in studies of YIG films grown on Gd$_3$Ga$_5$O$_{12}$ (GGG) substrates, where high-temperature growth or annealing often induces the formation of thin Gd-rich interfacial layers, identified as GdIG or Gd-doped YIG \cite{GomezPerez2018,Kumar2021PRB}. These interfacial layers couple antiferromagnetically to YIG and can give rise to pronounced exchange bias effects, inverted hysteresis loops, and strong suppression of the net magnetization \cite{Kumar,Roos2022}. Such observations underscore that interfacial magnetism in garnet systems is intrinsic and frequently dominates the static magnetic response.

%Exchange bias, manifested as a horizontal shift of the magnetic hysteresis loop, is a hallmark of interfacial exchange coupling and plays a critical role in stabilizing magnetic configurations in layered systems \cite{Nogues1999,Kiwi2001}. While traditionally associated with ferromagnet/antiferromagnet interfaces, exchange bias has also been observed in ferrimagnet/ferrimagnet systems, including garnet-based heterostructures \cite{Popova2003,Kumar2021PRB,Thuwal2025}. In YIG/GdIG systems, the compensated nature of GdIG introduces additional complexity, leading to temperature-dependent and sometimes unconventional bias behavior.

Given this strong interface sensitivity, a rigorous assessment of structural quality is a prerequisite for understanding the behavior of YIG/GdIG bilayers. High-resolution x-ray diffraction and transmission electron microscopy (TEM) are essential to verify epitaxy, crystallinity, and interface sharpness, as well as to identify possible interfacial phases or compositional gradients \cite{gomez2018synthetic, Roos2022}.

%While recent studies on YIG/GdIG heterostructures have highlighted rich dynamic phenomena such as ferromagnetic resonance, spin pumping, and thermally driven spin currents \cite{Fuhrmann2023,Becker2021}, these effects critically depend on the underlying static magnetic configuration and interface quality. Establishing a robust correlation between growth conditions, structural integrity, and static magnetism is therefore essential before addressing dynamical or transport phenomena.

In this work, we report the growth of epitaxial YIG/GdIG bilayer thin films and present a comprehensive study of their structural properties.
%By focusing on growth-induced effects and static magnetism, this study provides a reliable materials foundation for understanding garnet heterostructures and paves the way for future investigations of their dynamic properties.

\begin{table}[ht]
\centering
\caption{Sample nomenclature and corresponding layer structures investigated in this work. The multilayer stacks are listed from the top layer to the substrate.}
\label{tab:samples}
\begin{tabular}{|c|c|}
\hline
\textbf{Sample} & \textbf{Layer structure (top $\rightarrow$ substrate)} \\
\hline
SY  & Pt/YIG/GGG \\
\hline
SG  & Pt/GdIG/GGG \\
\hline
SYG & Pt/GdIG/YIG/GGG \\
\hline
SGY & Pt/YIG/GdIG/GGG \\
\hline
\end{tabular}
\end{table}

\section{Experimental Procedure}

\begin{figure}
\includegraphics[scale=0.15]{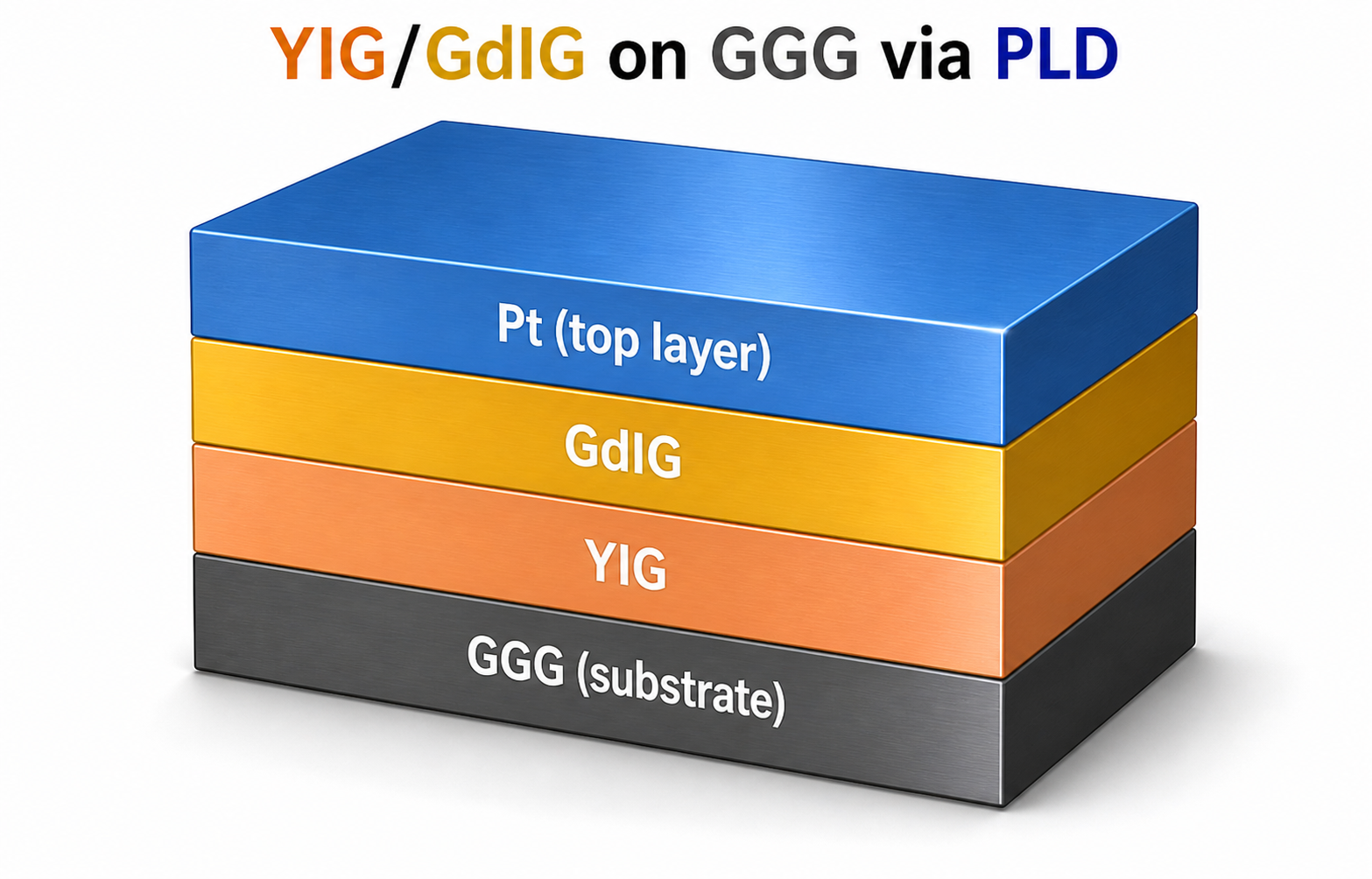}
\caption{\label{fig: Schematic} Schematic illustration of the sample structure SYG.}
\end{figure}

Epitaxial YIG ($\sim150$ nm) and GdIG ($\sim90$ nm) thin films were grown on (111)-oriented single-crystal Gd$_3$Ga$_5$O$_{12}$ (GGG) substrates using pulsed laser deposition (PLD) with a KrF excimer laser ($\lambda = 248$ nm). Prior to deposition, the growth chamber was evacuated to a base pressure of $\sim 3 \times 10^{-5}$ Pa. Stoichiometric targets of YIG and GdIG were used for depositing these thin films, and the target-to-substrate distance was fixed at 55 mm for all depositions. Both YIG and GdIG layers were deposited under identical growth conditions to ensure comparable crystalline quality and a well-defined interface. The substrate temperature during deposition was maintained at $750\,^{\circ}\mathrm{C}$ for both layers. The laser repetition rate was set to 5 Hz with a pulse energy of approximately 100 mJ, corresponding to an energy density of $\sim 1$ J/cm$^{2}$ on the target surface. The oxygen partial pressure during deposition was maintained at 14 Pa by introducing high-purity O$_2$ into the chamber. The thickness of each layer was independently optimized by varying the total number of laser pulses. YIG/GdIG bilayers were fabricated by sequential deposition of the YIG layer followed by the GdIG layer without breaking vacuum, ensuring minimal surface contamination and a clean interface. In the same deposition run, a Pt (10nm) layer was selectively deposited on part of the sample by masking the substrate, also using pulsed laser deposition, which will predominantly be discussed in our work. After deposition, the samples were cooled to room temperature at a controlled rate of 15\,$^\circ$C/min under the same oxygen ambient to preserve oxygen stoichiometry and reduce defect formation. Using this deposition procedure, single-layer YIG, single-layer GdIG, and bilayer YIG/GdIG heterostructures (as illustrated in Figure~\ref{fig: Schematic}), along with the corresponding reverse-order structures, were fabricated on GGG substrates, and the samples were named as listed in Table~\ref{tab:samples}.

The X-ray diffraction (XRD) measurements were performed using a Malvern Panalytical X'Pert Pro diffractometer equipped with a four-circle goniometer and a Cu K$\alpha_1$ radiation source ($\lambda = 1.5406$~\AA). The incident beam optics consisted of a parabolic mirror followed by a double Ge(220) hybrid monochromator, providing a highly monochromatic and collimated X-ray beam. On the diffracted beam side, no additional optics were employed, and the diffracted intensity was collected using a 255-channel PIXcel detector operating in one-dimensional (1D) mode. Symmetric $\theta$--2$\theta$ scans were employed to confirm phase purity while RSM measurements around asymmetric reflections provided detailed insight into the in-plane lattice coherence, strain state, and epitaxial relationship between the YIG/GdIG layers and the GGG substrate. To further examine the microstructural quality and interface sharpness of the heterostructures, cross-sectional transmission electron microscopy (TEM) was performed. Focused ion beam (FIB) lamellae were prepared using a Thermo Fisher Scientific Helios 5 UX system. High-resolution TEM (HRTEM) imaging, carried out using a FEI Tecnai G2 F20 S-TWIN HR(S)TEM microscope, enabled direct visualization of atomic lattice planes across the interfaces, while Fast Fourier transform (FFT)) patterns were used to confirm the crystallographic alignment of the multilayer stack. In addition, energy-dispersive x-ray spectroscopy (EDX) in scanning TEM mode was used to confirm the compositional distribution across the layers. 
%Complementary Raman spectroscopy measurements were carried out using a 532~nm excitation laser to probe the vibrational modes of the garnet films and to assess the structural integrity of the heterostructures (see Supplementary Fig.~S2 for details). The magnetic properties were investigated using a SQUID magnetometer, including temperature-dependent magnetization measurements to understand the evolution of the magnetic response and the influence of the interface in the YIG/GdIG heterostructures.

\section{Results and Discussion}

\subsection{Structural analysis: X-ray diffraction (XRD) and Reciprocal space mapping (RSM)}

\begin{figure*}
\includegraphics[scale=0.9]{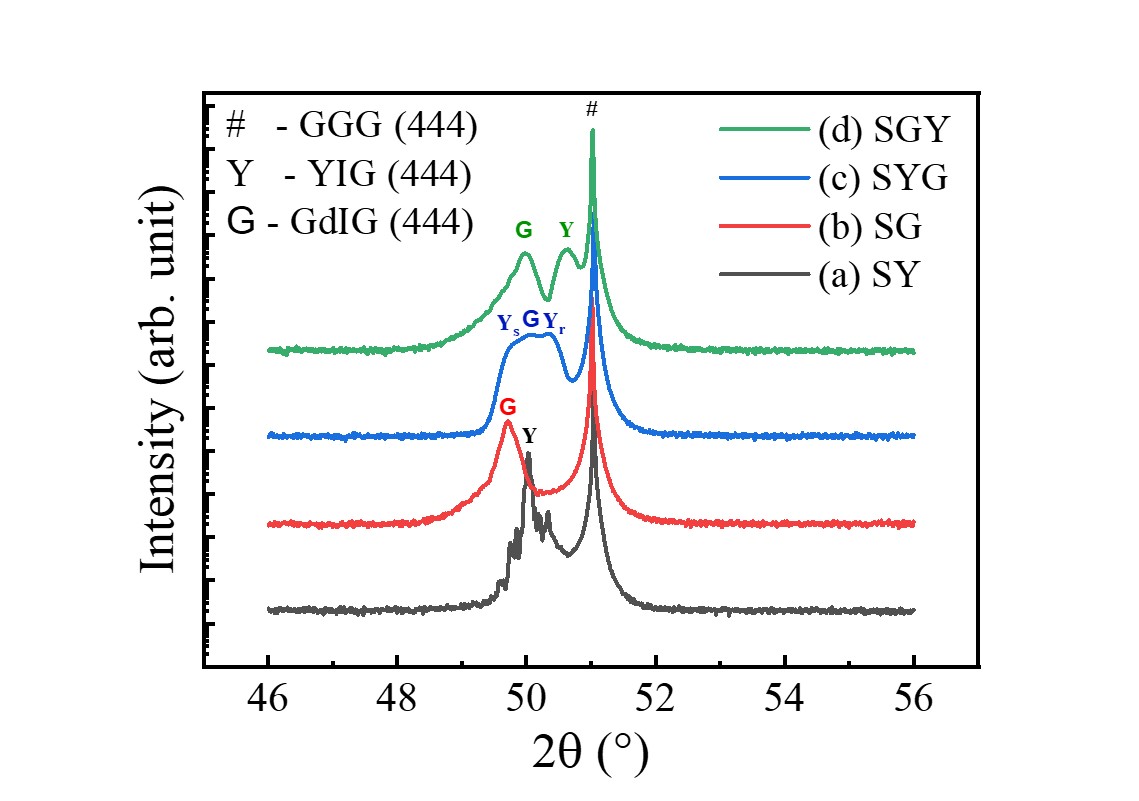}
6\caption{\label{fig: structural} XRD $\theta$–2$\theta$ scans of the deposited samples labeled as (a) SY, (b) SG, (c) SYG, and (d) SGY.}
\end{figure*}

\begin{figure*}
\includegraphics[scale=0.90]{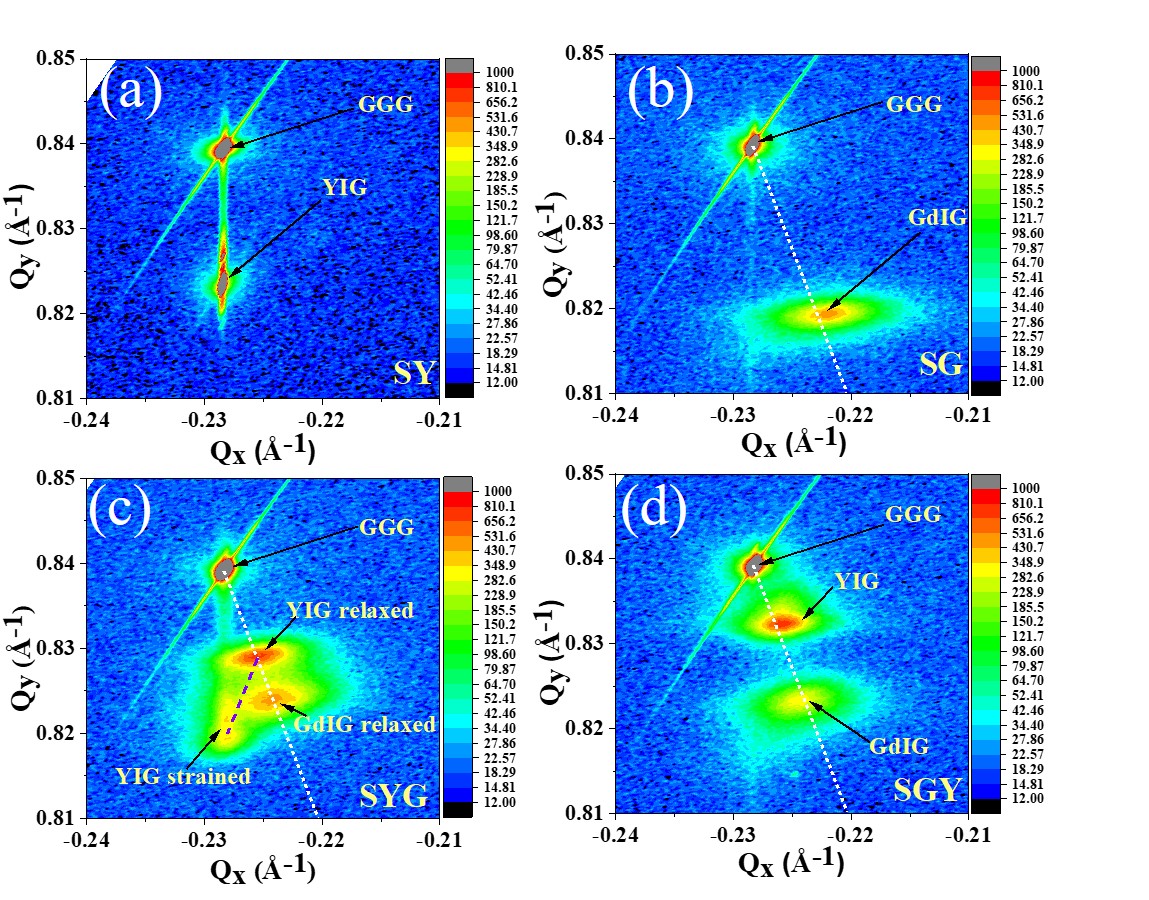}
\caption{ Reciprocal space maps around the asymmetric (486) reflection for (a) SY, (b) SG, (c) SYG, and (d) SGY. The white dotted line denotes the reciprocal-space positions expected for cubic structures with different lattice parameters, while the violet dashed line illustrates the strain evolution between the fully strained and relaxed YIG components.}
\label{fig:RSM_main}
\end{figure*}

Figure~\ref{fig: structural} shows the $\theta$--$2\theta$ XRD scans of samples
(a) SY, (b) SG, (c) SYG, and (d) SGY, and Figure~\ref{fig:RSM_main} shows the corresponding
RSMs, recorded around the asymmetric (486) reflection of the GGG substrate.
All samples exhibit a clear diffraction peak corresponding to the (444) reflection of the GGG
substrate, confirming that the films are grown along the (111) crystallographic direction,
consistent with the substrate orientation. To interpret the RSMs, two guide-to-the-eye lines are
included in Figure~\ref{fig:RSM_main}: a white dotted line connecting the reciprocal-space origin
with the GGG(486) substrate reflection, marking the locus of positions expected for cubic
structures of different lattice parameter, and a violet dashed line connecting the fully strained
and relaxed YIG reflections, illustrating the strain evolution from the coherently strained state
toward the relaxed cubic lattice.

To quantify the strain state, the in-plane and out-of-plane interplanar spacings were obtained
from the reciprocal-space coordinates of the (486) reflection: the out-of-plane spacing
$d_{444}$ is calculated as $2/3$ of the vertical (Q$_y$) projection, and the in-plane spacing
$d_{20\bar{2}}$ is the horizontal (Q$_x$) projection, following the RSM strain methodology of
Kubota~\textit{et al.}~\cite{kubota2013systematic}. For a cubic crystal grown along (111), the
in-plane biaxial strain $\varepsilon_{\parallel}$ and out-of-plane uniaxial strain
$\varepsilon_{\perp}$ are related through the elastic properties of the material via
\begin{equation}
\varepsilon_{\perp} = -\frac{2\nu}{1-\nu}\,\varepsilon_{\parallel},
\qquad\text{equivalently}\qquad
\nu = \frac{\varepsilon_{\perp}}{\varepsilon_{\perp}-2\varepsilon_{\parallel}},
\label{eq:poisson}
\end{equation}
which we use below to extract the Poisson's ratio ($\nu$) wherever both strain components of a
genuinely strained (non-reference) layer are available.

Rather than referencing the strain of each layer to the nominal, literature bulk lattice
parameters of YIG and GdIG, we take the relaxed YIG and GdIG components measured in the SYG
bilayer (indicated with $^{*}$ in Table~\ref{tab:strain}) as the reference structures, since these
correspond to the actual films grown in this work rather than to idealized bulk crystals. These
reference structures correspond to a cubic cell parameter $a=12.536$~\AA\ (YIG) and
$a=12.613$~\AA\ (GdIG) respectively $1.3\%$ and $1.1\%$ larger than the bulk values reported by
Geller and Gilleo~\cite{Geller1963}. Such lattice expansion is frequently reported for
PLD-grown garnet thin films and is commonly attributed to point defects introduced during growth,
including oxygen non-stoichiometry, Fe deficiency, or antisite
disorder~\cite{ning2021antisite,santiso2023antisite,Hauser2016,popova2003exchange}. Assuming for
the GGG substrate a cubic cell parameter $a=12.379$~\AA\ (PDF: 01-083-4878), i.e.\
$d_{20\bar{2}}=4.377$~\AA, the in-plane mismatch for YIG and GdIG grown directly on GGG is
$-1.3\%$ and $-1.8\%$, respectively, while GdIG grown on YIG is $-0.52\%$ and, in the inverted
sequence, YIG grown on GdIG is $+0.52\%$. All strain and mismatch values discussed below are
referenced to these actual, relaxed film structures rather than to ideal bulk crystals.
Below, we present a sample-by-sample comparison of all the heterostructures based on the XRD data obtained as shown in Figure~\ref{fig: structural} and Figure~ \ref{fig:RSM_main}:

\paragraph{SY (single-layer YIG):}
The YIG(444) diffraction peak is observed adjacent to the GGG(444) substrate peak, reflecting the
small lattice mismatch between YIG and GGG. The peak shows a structured profile consisting of
multiple closely spaced maxima superimposed with weak thickness fringes, suggesting the
coexistence of regions with slightly different out-of-plane lattice parameters rather than a
single, perfectly homogeneous layer; the narrow linewidth of the individual maxima nonetheless
indicates high crystalline quality. This is supported by the RSM, which shows the YIG peak
vertically aligned with the GGG substrate peak, confirming that the in-plane lattice parameter of
the film closely matches that of the substrate and that the YIG layer remains fully strained.
The measured in-plane strain, $\varepsilon_{\parallel}=-1.31\%$, is in excellent agreement with the
mismatch expected for YIG grown directly on GGG, confirming that this layer is
essentially fully strained. The corresponding out-of-plane strain, $\varepsilon_{\perp}=+0.67\%$,
reflects the expected tetragonal-like (rhombohedral) elastic response to this in-plane
compression, corresponding to $d_{444}=1.8217$~\AA\ and $d_{20\bar{2}}=4.3783$~\AA. Using
Eq.~\eqref{eq:poisson}, this strain pair yields a Poisson's ratio $\nu\approx0.204$, a physically
reasonable value for a cubic garnet.

\begin{table*}
\centering
\caption{Structural parameters extracted from the asymmetric (486) reciprocal space maps. The out-of-plane ($d_{444}$) and in-plane ($d_{2\bar{2}0}$) interplanar spacings together with the corresponding out-of-plane ($\varepsilon_{\perp}$) and in-plane ($\varepsilon_{\parallel}$) strain components are listed for all samples. Values marked with an asterisk (*) correspond to the experimentally observed relaxed components, which were used as the reference lattice parameters for the strain calculations.}

\label{tab:strain}
\renewcommand{\arraystretch}{1.25}
\begin{tabular}{llcccc}
\toprule
\textbf{Sample} & \textbf{Layer} &
\textbf{$d_{444}$ (\AA)} &
\textbf{$d_{2\bar{2}0}$ (\AA)} &
\textbf{$\varepsilon_{\perp}$ (\%)} &
\textbf{$\varepsilon_{\parallel}$ (\%)} \\
\midrule

SY & YIG &
1.8217 &
4.3783 &
+0.67 &
$-$1.31 \\

\midrule

SG & GdIG &
1.8309 &
4.5057 &
+0.56 &
+1.03 \\

\midrule

\multirow{3}{*}{SYG}
& YIG (strained component)
& 1.8244
& 4.4364
& +1.22
& $-$0.96 \\

& YIG (relaxed component)
& 1.8094$^{*}$
& 4.4364$^{*}$
& 0
& 0 \\

& GdIG
& 1.8206$^{*}$
& 4.4597$^{*}$
& 0
& 0 \\

\midrule

\multirow{2}{*}{SGY}
& GdIG
& 1.8218
& 4.4561
& +0.07
& $-$0.08 \\

& YIG
& 1.8023
& 4.4271
& $-$0.39
& $-$0.20 \\

\bottomrule
\end{tabular}
\end{table*}

\paragraph{SG (single-layer GdIG):}
The GdIG(444) diffraction peak is comparatively broader than the YIG(444) peak of sample SY,
suggesting a reduced structural coherence length and/or increased lattice inhomogeneity, likely
arising from defects, strain fluctuations, mosaicity, or slight compositional variations
introduced during growth~\cite{ning2021antisite,santiso2023antisite,Hauser2016,popova2003exchange}.
The RSM shows the GdIG reflection broadened and shifted in $Q_x$ relative to the substrate peak,
indicating that the film is partially relaxed, with a faint residual intensity aligned with the
substrate peak suggesting a small, fully strained interfacial
component~\cite{popova2003exchange,gomez2018synthetic}. Quantitatively, however, the measured
in-plane strain is $\varepsilon_{\parallel}=+1.03\%$ ($d_{20\bar{2}}=4.5057$~\AA), i.e.\ tensile,
which is not physically possible for a film grown on a substrate with a smaller cell parameter
(a compressive strain of about $-1.8\%$ would be expected). This indicates that the GdIG structure
in this sample differs from the reference structure, with a genuinely larger cell volume; the
positive out-of-plane strain, $\varepsilon_{\perp}=+0.56\%$ ($d_{444}=1.8309$~\AA), supports an
overall volume increase rather than a simple elastic response. Applying Eq.~\eqref{eq:poisson} to
this pair formally gives $\nu\approx-0.37$ -- an unphysical, negative value for a non-auxetic
garnet -- which is itself a quantitative confirmation that this strain pair does not represent a
simple elastic biaxial-strain response, consistent with the compositional-variation origin
proposed above.

\paragraph{SYG (YIG/GdIG bilayer):}
The XRD pattern shows a single broadened peak rather than two resolved YIG and GdIG reflections, a
consequence of the small lattice mismatch between the two garnets combined with the moderate film
thickness, which causes the individual reflections to overlap within the instrumental resolution;
this is consistent with the comparable interplanar spacings obtained from the corresponding
single-layer films. The RSM, however, resolves three distinct components. The YIG layer shows both
a strained and a relaxed component: the strained component lies along the violet dashed guide
line, with $\varepsilon_{\parallel}=-0.96\%$ and $\varepsilon_{\perp}=+1.22\%$
($d_{444}=1.8244$~\AA), values comparable to the single-layer SY case, with the somewhat larger
$\varepsilon_{\perp}$ likely combining the expected elastic response with a modest change in cell
volume of this YIG portion; applying Eq.~\eqref{eq:poisson} gives $\nu\approx0.39$. The relaxed
YIG component and the GdIG reflection both lie close to the cubic guide line and are, by
construction, taken as the reference structures for this work ($\varepsilon_{\parallel}=
\varepsilon_{\perp}=0$; $d_{444}=1.8094$~\AA, $d_{20\bar{2}}=4.4364$~\AA\ for YIG, and
$d_{444}=1.8206$~\AA, $d_{20\bar{2}}=4.4597$~\AA\ for GdIG), so no Poisson's ratio can be (or needs
to be) extracted for these two components. The coexistence of strained and relaxed YIG components
likely reflects strain redistribution during bilayer growth, while the GdIG layer is predominantly
relaxed, with only the small residual mismatch expected from its interfacial interaction with the
underlying YIG layer.

\paragraph{SGY (GdIG/YIG bilayer):}
In contrast to SYG, the XRD pattern of SGY shows resolved YIG(444) and GdIG(444) features, likely
reflecting differences in strain relaxation and interfacial quality associated with the reversed
growth sequence. The RSM shows that the GdIG layer, grown directly on GGG, undergoes substantial
relaxation rather than remaining fully strained: the measured strain,
$\varepsilon_{\parallel}=-0.08\%$ and $\varepsilon_{\perp}=+0.07\%$ ($d_{444}=1.8218$~\AA,
$d_{20\bar{2}}=4.4561$~\AA), is far smaller in magnitude than the $-1.8\%$ mismatch expected for a
coherently strained GdIG/GGG interface, confirming that the larger GdIG/GGG mismatch prevents
coherent epitaxial growth at this thickness and the layer relaxes almost completely; the very small
residual strain nonetheless yields a reasonable $\nu\approx0.30$ via Eq.~\eqref{eq:poisson},
consistent with the SY value, though this estimate should be treated cautiously given the small
strain magnitudes involved. The YIG layer subsequently grown on the relaxed GdIG shows
$\varepsilon_{\parallel}=-0.20\%$, which is counterintuitive: YIG grown on relaxed GdIG is expected
to be under tensile in-plane strain ($+0.52\%$ mismatch), not compressive. Together with the
out-of-plane strain, $\varepsilon_{\perp}=-0.39\%$ ($d_{444}=1.8023$~\AA,
$d_{20\bar{2}}=4.4271$~\AA) indicating an overall reduction in cell volume rather than an
elastic expansion, this indicates that the YIG cell parameters in this layer are governed by
composition variations rather than elastic strain alone. Indeed, formally applying
Eq.~\eqref{eq:poisson} to this pair is not meaningful: because $\varepsilon_{\perp}\approx
2\varepsilon_{\parallel}$, the denominator collapses toward zero and returns an arbitrarily large,
unphysical value for $\nu$, itself confirming that no simple elastic biaxial-strain description
applies to this layer. The change in growth order therefore significantly influences the strain
evolution and lattice relaxation behavior within the heterostructure.

It is also worth noting that the RSM measurements are highly reproducible across the sample
series, with no essential differences observed between samples prepared within the same
deposition run; the RSM patterns of Pt-capped and uncapped samples are nearly identical,
indicating that the structural properties are predominantly determined by the growth conditions.
Similar growth-order-dependent strain and relaxation
behavior has been reported in other garnet heterostructures~\cite{Roos2022,kumar2021positive}.

\subsection{Cross-sectional Transmission Electron Microscopy analysis}

\begin{figure*}
\includegraphics[scale=0.6]{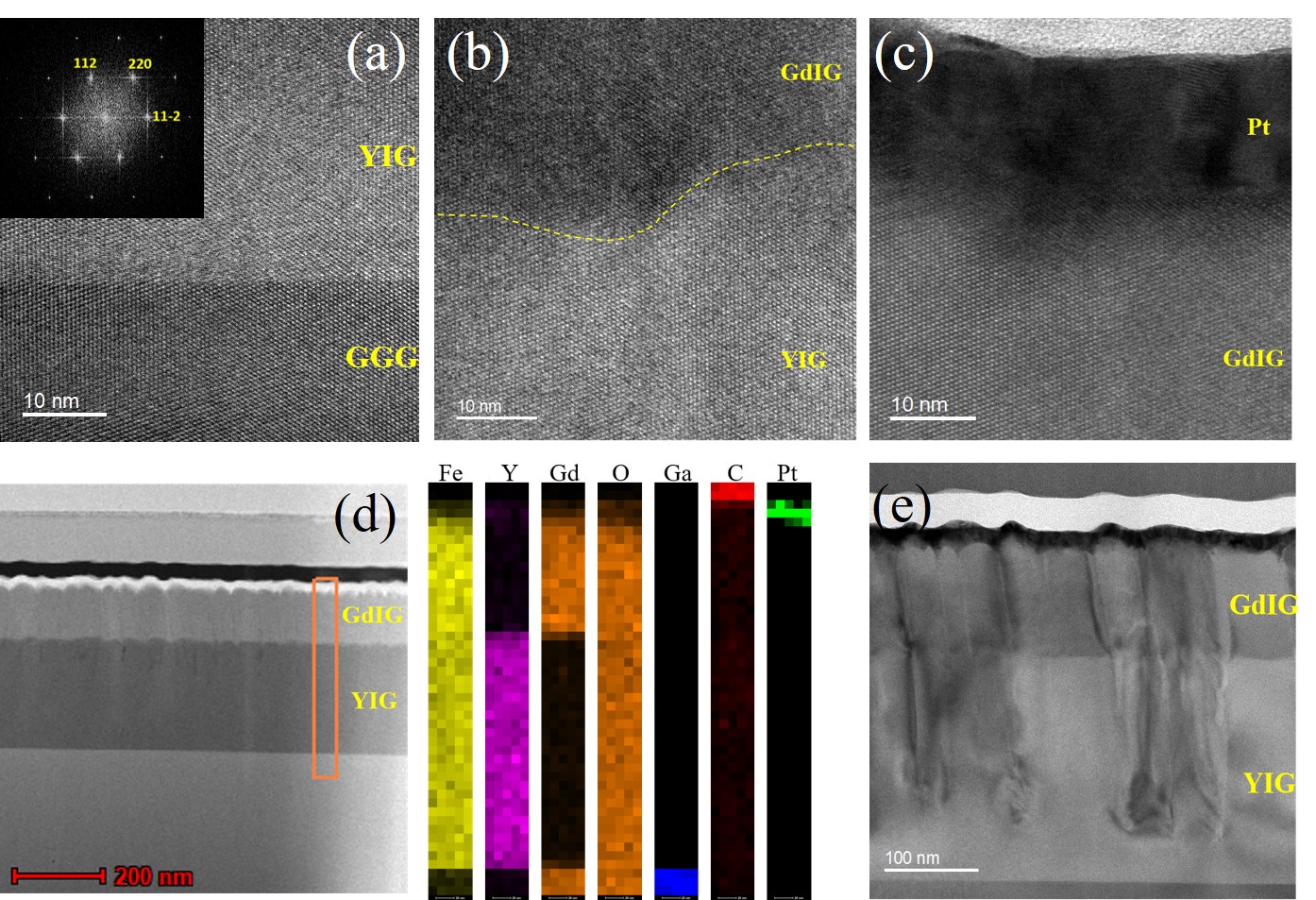}
\caption{\label{fig: TEM_YG} Cross-sectional TEM images of sample SYG. (a) GGG/YIG interface with the corresponding FFT pattern (inset). (b) YIG/GdIG interface. (c) GdIG/Pt interface. (d) Low-magnification STEM image of the full GGG/YIG/GdIG/Pt stack with corresponding EDX elemental mapping. (e) Magnified view of the multilayer stack}
\end{figure*}

\begin{figure*}
\includegraphics[scale=0.7]{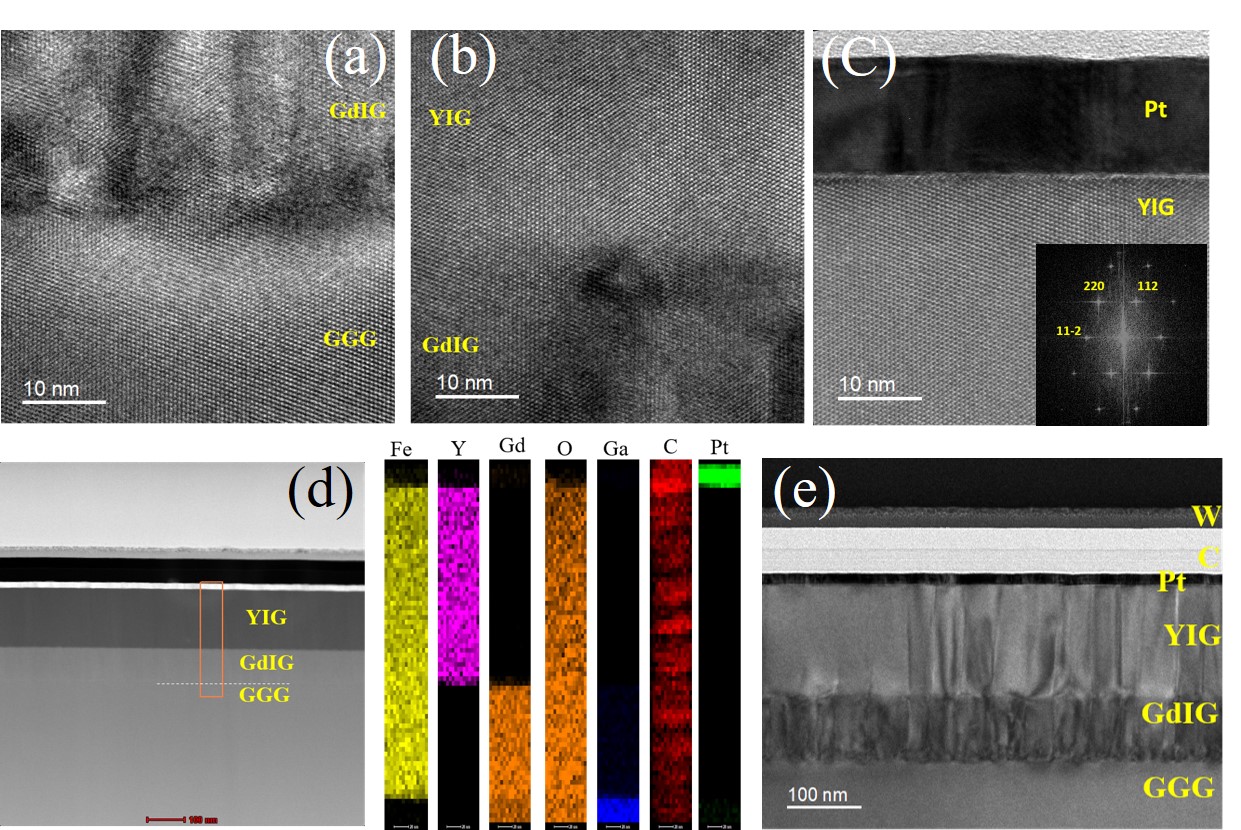}
\caption{\label{fig: TEM_GY} Cross-sectional TEM images of sample SGY. (a) GGG/GdIG interface.
(b) GdIG/YIG interface. (c) YIG/Pt interface with the corresponding FFT pattern shown in the
inset. (d) Low-magnification STEM image of the complete GGG/GdIG/YIG/Pt stack with EDX elemental
mapping. (e) Magnified view of the multilayer stack.}
\end{figure*}

\begin{figure*}
\includegraphics[scale=0.6]{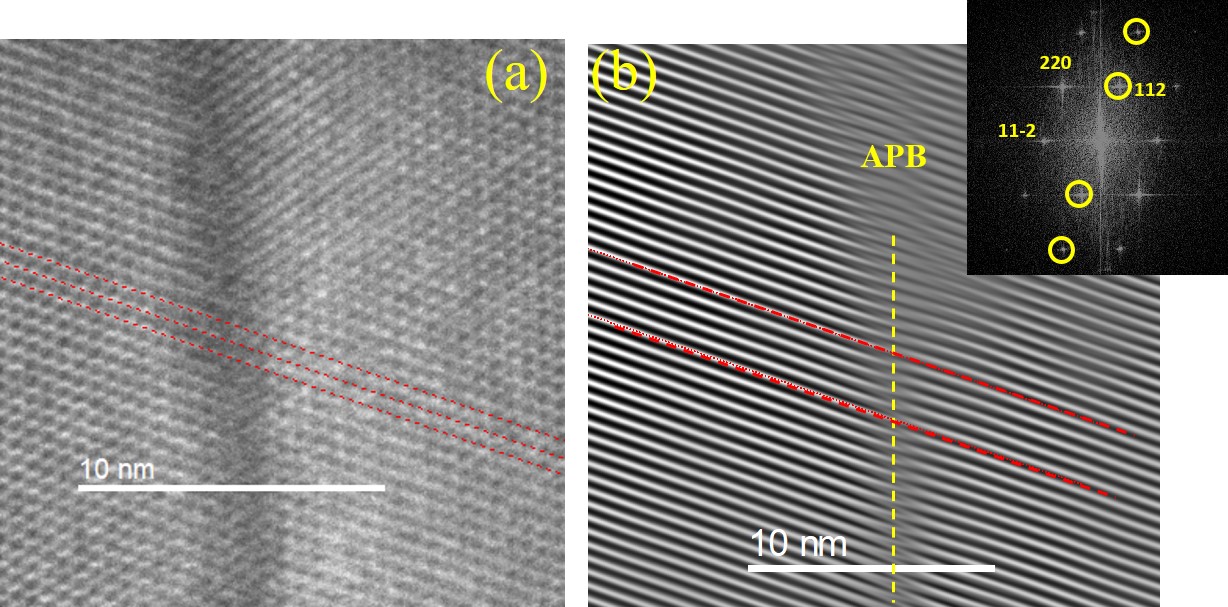}
\caption{\label{fig: APB} TEM analysis of antiphase boundaries (APBs) in the sample SGY. (a) HRTEM image showing APBs. (b) Inverse FFT (IFFT) highlighting the APB region (inset). (c) Corresponding FFT of the selected area.}
\end{figure*}

Cross-sectional TEM was performed to further investigate the microstructural quality and interfacial structure of the YIG/GdIG heterostructures. TEM lamellae of samples SYG and SGY were prepared using FIB milling technique. Prior to ion milling, a protective carbon layer followed by a tungsten layer was deposited on the sample surface in order to minimize surface damage and prevent ion-induced amorphization during the FIB thinning process. Such protective coatings are commonly used in FIB sample preparation to preserve the integrity of the film interfaces and ensure reliable high-resolution imaging \cite{WilliamsCarter2009}. The lamellae were subsequently thinned and transferred to a TEM grid for cross-sectional imaging. The cut orientation was carefully selected to align along the [10$\bar{1}$] zone axis, which is a commonly used crystallographic direction for imaging garnet structures.

Figure~\ref{fig: TEM_YG} shows the cross-sectional TEM characterization of sample SYG. The interface between the GGG substrate and the YIG layer, shown in Figure~\ref{fig: TEM_YG}(a), appears sharp and well-defined, indicating high-quality epitaxial growth. The corresponding Fast Fourier transform (FFT) pattern displayed in the inset confirms the crystallographic alignment between the film and the substrate, consistent with the epitaxial relationship inferred from the XRD and RSM measurements. High-resolution imaging of the YIG layer reveals that the initial growth region, approximately the first 50–60 nm above the substrate, is nearly defect-free and exhibits excellent lattice coherence with the underlying GGG substrate.

Beyond this thickness, however, structural defects begin to appear within the YIG layer, giving rise to a columnar microstructure that propagates through the film thickness. These defects become particularly evident in the low-magnification TEM images shown in Figure~\ref{fig: TEM_YG}(d), where strain fields and columnar domains can be clearly distinguished. Such a microstructural evolution may explain the presence of two different mechanical states in YIG crystal observed in the XRD measurements, corresponding to strained and relaxed components within the film.

The interface between the YIG and GdIG layers, shown in Figure~\ref{fig: TEM_YG}(b), is slightly wavy but remains well-defined. The contrast observed across the interface does not indicate significant interdiffusion between Y and Gd atoms, suggesting that the interface remains chemically sharp. This observation is further supported by the EDX elemental mapping shown in Figure~\ref{fig: TEM_YG}(d), which confirms the spatial separation of Y and Gd across the heterostructure. The columnar structure originating in the upper region of the YIG layer appears to propagate into the GdIG layer, indicating that the microstructural features of the underlying layer influence the subsequent growth of the overlayer. Such columnar growth is commonly observed in oxide thin films deposited by PLD when strain relaxation occurs through defect formation \cite{Hauser2016}.

The interface between GdIG and the Pt layer is shown in Figure~\ref{fig: TEM_YG}(c). Additional insight into the microstructure is obtained from the low-magnification TEM image shown in Figure~\ref{fig: TEM_YG}(e), where the entire multilayer stack is visible. The image reveals alternating columns within the YIG layer that exhibit different strain states. Some columns appear to remain coherent with the bottom YIG region, while others show larger strain contrast. An interesting observation is that the columnar domains appear tilted in a common direction, suggesting anisotropic strain relaxation within the film \cite{allen1979microscopic}.

Figure~\ref{fig: TEM_GY} shows the TEM characterization of the inverted bilayer sample SGY (GGG/GdIG/YIG/Pt). In this structure, the GdIG layer is deposited directly on the GGG substrate. The interface between GdIG and GGG, shown in Figure~\ref{fig: TEM_GY}(a), exhibits a relatively high density of defects compared to the YIG/GGG interface observed in SYG. This indicates that GdIG nucleates with a higher defect density when grown directly on the GGG substrate. Nevertheless, the interface remains sharp and no significant interdiffusion between Y and Gd is observed.

Interestingly, the YIG layer deposited on top of GdIG exhibits improved crystalline quality. As shown in Figure~\ref{fig: TEM_GY}(b), the YIG/GdIG interface appears relatively smooth and well defined. In several regions, the defects present in the underlying GdIG layer propagate into the YIG film, whereas in other areas the YIG layer grows with significantly fewer defects. This suggests that the defect propagation is not uniform across the film. The surface of the YIG layer appears remarkably flat compared to the YIG layer in SYG.

The interface between the YIG layer and the Pt layer, shown in Figure~\ref{fig: TEM_GY}(c), appears well-defined and smooth, indicating good surface quality of the top YIG layer as compared to that of GdIG layer in Figure~\ref{fig: TEM_YG}(c). The FFT pattern in the inset confirms the crystallinity of the underlying garnet structure. EDX elemental maps in Figure~\ref{fig: TEM_GY}(d) clearly distinguish the different layers and show no significant interdiffusion between Y and Gd across the interfaces. Minor contrast variations suggest a relatively higher Fe signal in the GdIG layer compared to YIG, along with a slightly enhanced Gd contrast in GdIG relative to the GGG substrate, consistent with their compositional differences.

A closer examination of the microstructure of the SGY sample reveals the presence of antiphase boundaries (APBs) within the YIG layer, as illustrated in Fig.~\ref{fig: APB}. Such defects are commonly observed in epitaxial garnet films and originate from the coalescence of independently nucleated domains with identical crystallographic orientation but different translational \cite{Geller1963}. In the HRTEM image shown in Fig.~\ref{fig: APB}(a), a boundary region can be clearly identified where the lattice planes exhibit a relative translation. On one side of the boundary, the bright atomic columns associated with the ${112}$ lattice planes are aligned along the parallel lines indicated by the red dashed lines, whereas on the opposite side the continuation of these lines intersects the dark atomic columns of the same family of planes. This contrast inversion corresponds to a relative displacement of approximately half an interplanar spacing, representing the characteristic signature of an antiphase boundary \cite{wu2016anti,celotto2003characterization}. Such a translation introduces a phase shift in the periodic lattice while preserving the overall crystallographic orientation.

The APB becomes more clearly resolved after selectively filtering the ${112}$ reflections from the fast Fourier transform (FFT) of the HRTEM image. The corresponding inverse FFT (IFFT) reconstruction retains only the periodic lattice information associated with the ${112}$ planes, thereby enhancing the phase contrast across the defect. As shown in Fig.~\ref{fig: APB}(b) and (c), this procedure reveals a distinct phase discontinuity manifested as a lateral displacement of the lattice fringes.

Importantly, APBs represent crystallographic defects that locally disrupt the structural coherence of the epitaxial lattice. In garnet systems, where magnetic ordering is governed by Fe–O–Fe superexchange interactions, the local distortion of bond lengths and bond angles across the APB modifies the exchange pathways and may induce localized spin canting, magnetic inhomogeneity, and a reduction of the local magnetic moment \cite{wu2016anti,Cherepanov1993}. The observation of APBs in the SGY heterostructure therefore indicates a comparatively higher defect density associated with this growth sequence. In contrast, the SYG heterostructure exhibits a sharper interface and does not show evidence of such extended planar defects, suggesting superior structural coherence and making it a more suitable architecture for subsequent spintronic and magnonic applications. \cite{wu2016anti, celotto2003characterization,Cherepanov1993}

%The corresponding FFT pattern used for the filtering is presented in Figure~\ref{fig: APB}(c). From a crystallographic standpoint, the APB represents a planar defect characterized by a non-lattice translation vector, which introduces localized lattice distortions and strain fields. These regions act as effective strain relaxation centers in epitaxial films and are often correlated with the onset of columnar growth observed in thicker layers.

\section{Conclusion}

In summary, we demonstrate the successful realization of epitaxial YIG/GdIG bilayer heterostructures on GGG(111) substrates and systematically investigate their structural properties. The films exhibit high crystalline quality with well-defined and coherent interfaces, as confirmed by XRD, RSM, and TEM analyses. The strain state of the heterostructures shows the coexistence of strained and partially relaxed regions, governed primarily by the growth sequence and interfacial lattice matching. Notably, the two stacking configurations exhibit distinct interfacial characteristics, reflected in differences in interface sharpness, microstructural features, and strain distribution across the multilayer stack.

A comparative analysis reveals that YIG grown directly on the GGG substrate exhibits superior crystalline quality with reduced defect density and a more uniform microstructure, which facilitates more coherent growth of the subsequent GdIG layer in the GGG/YIG/GdIG configuration. In contrast, GdIG grown directly on the substrate shows slightly higher interfacial defect density, which partially propagates into the overlying YIG layer and may influence the magnetic and spintronic properties of these heterostructures. These results indicate that the YIG-first stacking sequence is structurally more favorable, offering improved strain relaxation and interface quality, and highlight the importance of growth sequence in tailoring garnet heterostructures. The relatively larger film thicknesses employed in this work are intentionally chosen to enable a clearer investigation of magnon coupling dynamics and their impact on spintronic parameters such as torque efficiency, as well as to ensure sufficiently strong signals for future studies of interfacial exchange and spin transport phenomena in these systems for spintronic and magnonic applications.

\section{Acknowledgement}
S.B., A.S., K.S.R., P.G., S.P.M, S.S., and A.M. thank the Department of Atomic Energy (No. 0803/2/2020/NISER/R\&D-II/8149), Department of Science and Technology, Science and Engineering Research Board (Grant No. CRG/2021/001245) for providing financial support. S.B., A.M. also acknowledges funding support for Chanakya Postdoctoral fellowship from the National Mission on Interdisciplinary Cyber Physical Systems, of the Department of Science and Technology, Govt. of India through the IHUB Quantum Technology Foundation of India, (Sanction Order No. I-HUB/PDF/2022-23/04). All the authors thank the funding from the European Union’s Horizon 2020 research and innovation programme under grant agreement No 101007417, having benefited from the access provided by ICN2 in UAB within the framework of the NFFA-Europe Pilot Transnational Access Activity, proposal ID462.

\bibliography{Manuscript}% Produces the bibliography via BibTeX.
\end{document}